\documentclass{article}
\usepackage[tbtags]{amsmath}
\usepackage{amsfonts,verbatim,amssymb}
\usepackage{mathrsfs}
\usepackage{showlabels}
\usepackage[paper]{mn2e}   %mn2e.sty + mnpaper.sty required
\usepackage{float}
\allowdisplaybreaks

\def\Russpages{00--00}

\theoremstyle{plain}
\newtheorem{theorem}{Theorem}

\theoremstyle{definition}

\newtheorem{demo}{Proof}

\def\qed{\quad\text{$\square$}}

\begin{document}
\hyphenation{dom-i-na-tion con-sid-ered}

\date{%
Received March~1, 2015%
%; in final form, May~1?, 20??%
}

\author{D.~DOGAN DURGUN}
\address{Manisa Celal Bayar University, Martyr Prof. Dr. Ilhan Varank Campus, Manisa, Turkey}
\email{derya.dogan@cbu.edu.tr}

\author{ E.\ N.~TOPRAKKAYA}
\address{Manisa Celal Bayar University, Martyr Prof. Dr. Ilhan Varank Campus, Manisa, Turkey}
\email{toprakkayaemre@gmail.com}

%%HINT: Insert ALL the author(s) below for future use,
%%preserving the percentage signs!!
%\authorinfo%page%
%{I.~I.~Ivanov, P.~P.~Petrov, and S.~S.~Sidorov}

%%TITLE LINES OF NO MORE THAN 55 SYMBOLS
\title{%
Roman Domination \\
of the Comet, Double Comet, and Comb Graphs}

\maketitle

%%INSERT THE RUNNING HEADS BELOW (NO INITIALS, 65 symbols at most!!)
%\markboth{Ivanov, Petrov??}{Short Title??}
%HINT: For more than two authors:
\markboth{D.~DOGAN DURGUN, E.\ N.~TOPRAKKAYA}%
{Roman Domination of the Comet, Double Comet, and Comb Graphs}

\begin{abstract} %%Not needed for short communications
 One of the well-known measurements of vulnerability in graph theory is domination.
There are many kinds of dominating and relative types of sets in graphs. However, we are going to focus on Roman domination, which is a type of domination that has historical and mathematical origins. The Roman domination numbers of the comet, double comet, and comb graphs are given in this paper.
\end{abstract}

%%The first keyword must begin with a small letter
%%(unless it requires a capital letter)
\begin{keywords}
Graph Theory, Vulnerability, Domination, Roman Domination.
\end{keywords}
%-----------------------------------------------

\section{Introduction}
%\label{s1}
A graph is a pair of sets $G = (V, E)$, where $V$ is the set of vertices and $E$ is the set of edges, formed by pairs of vertices.
 An area of graph theory that has received attention during recent decades is that of domination in graphs.
 A vertex $v$ in a graph $G$ is said to dominate itself and each of its neighbors, that is, $v$ dominates the vertices in its closed neighborhood $N[v]=\{u \in V:uv \in E\}\cup \{v\}$.
  A set $S$ of vertices of $G$ is a dominating set of $G$ if every vertex of $G$ is dominated by at least one vertex of $S$.
   Equivalently, a set $S$ of vertices of $G$ is a dominating set if every vertex in $V(G) - S$ is adjacent to at least one vertex in $S$.
    The minimum cardinality among the dominating sets of $G$ is called the domination number of $G$ and is denoted by $\gamma(G)$.
     A dominating set of cardinality $\gamma(G)$ is then referred to as a minimum dominating set. \par Now consider a military unit.
      Each military unit from the largest to the smallest has a very clear chain of command. Therefore, there must be free-flowing communication between the commanding echelon and soldiers.
        For easy commandment, each of the soldiers should be under command of at least one commander. For example, consider a battalion's graph model $G$.
         Let each person in the battalion be a vertex. If a soldier and a commander are linked to each other by commandment relation, connect these two vertices with an edge.
          In the battalions with this characteristic property, a selected set of soldiers $S$ or a selected set of commanders $V(G)-S$ are dominating sets of this battalion's graph model.
           \par We study a variant of domination, called roman domination, that came up with an article by Ian Stewart ~\cite{1},  of which origin is about the military strategy of the Roman Empire in the 4th century.
            Which is why we gave an example about military units before, for making the reader familiar with the subject. Let $G= (V, E)$ be a graph,
             the function $f: V \to \{0,1,2\}$ satisfying the condition that every vertex $u$ for which $f(u) = 0$ is adjacent to
              at least one vertex $v$ for which $f(v) = 2$ is a Roman dominating function (RDF). The weight of an RDF is the value $f(V)=\underset{u \in V}\sum{f(u)}$,
               which equals $|V_1|+2|V_2|$,  and the minimum weight of an RDF on a graph $G$ is called the Roman domination number of $G$ in ~\cite{2},
denoted by $\gamma_R(G)$. For an RDF $f$, let $V_i(f)=\{v \in V(G): f(v)=i\}$. In the context of a fixed RDF,
 we suppress the argument and simply write $V_0, V_1$, and $V_2$.  Since this partition determines $f$, we can equivalently write $f=(V_0, V_1, V_2)$
~\cite{3}.
 A function $f=(V_0, V_1, V_2)$ is a $\gamma_R-function$ if it is an RDF and $f(V)=\gamma_R(G)$ ~\cite{2}.
 \par In this paper, the roman domination number of the comet, double comet, and comb graphs are generalized and given with their proofs.

 \section{Preliminary}
 For $t\geq2$ and $r\geq1$, \textit{the comet graph} $C_{t,r}$ with $t+r$ vertices is the graph obtained by identifying one end of the path $P_t$ with the center of the star $K_{1,r}$ ~\cite{4}.
\begin{figure}[H]
    \centering
    \includegraphics[width= 0.5\textwidth]{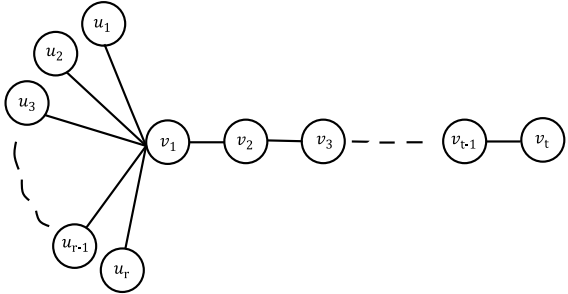}
    \caption{Comet Graph $C_{t,r}$}
    %\label{fig:Comet Graph}
\end{figure}
\par
For $a, b \geq 1$, $n\geq a+b+2$ by $DC(n, a, b)$ we denote a \textit{double comet graph}, which is a tree composed of a path containing $n-a-b$ vertices with $a$ pendent vertices attached to one of the ends of the path and $b$ pendent vertices attached to the other end of the path ~\cite{5}.
\begin{figure}[H]
    \centering
    \includegraphics[width= 0.5 \textwidth]{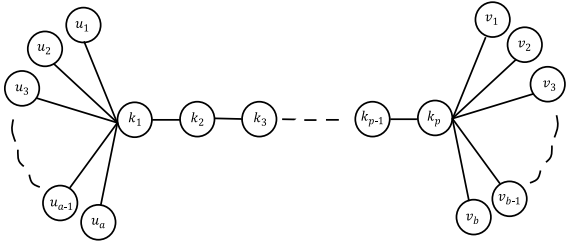}
    \caption{Double Comet Graph $DC(n, a, b)$}
    %\label{fig:Double Comet Graph}
\end{figure}
\par
A vertex of a graph is said to be \textit{pendent} if its neighborhood contains exactly one vertex. An edge of a graph is said to be pendent if one of its vertices is a pendent vertex.
The comb graph $P_n\mathrm{\Theta}\ K_1$, is the graph obtained from a path $P_n$ by attaching pendent edge at each vertex of the path and is denoted by $P_n^+$ ~\cite{6}.
\begin{figure}[ht]
    \centering
    \includegraphics[width=0.5 \textwidth]{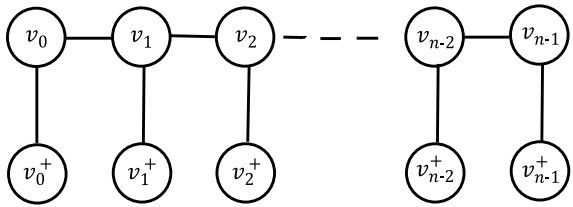}
    \caption{Comb Graph $P_n^+$}
    %\label{fig:Comb Graph}
\end{figure}

\section{Main Results}
In this section, we shall give the roman domination numbers of three different graphs. The proof of the roman domination number for comb graph, we use $v^+_t$  for the pendent vertices of $v_t$.
\begin{theorem}
Let $G=C_{t,r}$ be a comet graph where $t\geq2$ and $r\geq1$. Then the roman domination number of $G$ is equal to
\begin{equation}
{\gamma}_R(C_{t,r})=
\begin{cases}
2\frac{t}{3}+1 & t  {\equiv}0  \text{(mod 3)}\\
2\lceil \frac{t}{3} \rceil &  otherwise
\end{cases}
\end{equation}
\end{theorem}
\begin{demo}
Roman domination number of the comet graph is considered in three cases. Let $f=(V_0, V_1, V_2)$ be a $\gamma_R-function$ of $G$. \bigskip\par
(1) $t {\equiv} 0 \text{(mod 3)}$\par
In order to dominate $u_1, u_2, u_3, \ldots, u_r$ , $v_1$ and $v_2$ vertices, $v_1$ vertex should be taken into $V_2$ set. To dominate $v_{t-2}$, $v_{t-1}$ and $v_{t-3}$ vertices, $v_{t-2}$ vertex should be taken into $V_2$ set. To dominate $v_{t}$ vertex, $v_{t}$ itself should be taken into $V_1$ set. For the rest vertices of the graph which are not dominated, $v_t$ vertices should be taken into $V_2$ set which satisfy $t \equiv 1 \text{(mod 3)}$. Then for a Roman dominating function (RDF) $f$, $V_2=\lbrace v_1, v_4, v_7, \ldots, v_{t-2}\rbrace$ and $V_1=\lbrace v_{t}\rbrace$ .\par
So that $f(V)=1+2(\frac{t-2-1}{3}+1)$ then we get $\gamma_R(C_{t,r})\leq 2\frac{t}{3}+1$. \par
Let f not be a $\gamma_R-function$ and by deleting $v_t$ vertex from $V_1=\{v\in V: f(v)=1\}$ set, let $V_1=\emptyset$. Since $f'(v)\neq 2$ for $\forall v \in N(v_t)$, obtained function $f'=(V_0 \cup \{v_t\},\emptyset, V_2)$ does not satisfy the condition to be an RDF. According to this $\gamma_R(C_{t,r})\geq 2\frac{t}{3}+1$. For $f'$ function to be an RDF, $v_t$ vertex should be taken into $V_2$ set; $f'=(V_0,\emptyset,V_2\cup\{v_t\})$. Hence we get $f'(V)=2\frac{t}{3}+2$. Since $f(V)<f'(V)$ that $f'(V)\neq\gamma_R(G)$. In this case $\gamma_R(C_{t,r})\geq2\frac{t}{3}+1$.\par
Let f not be a $\gamma_R-function$ and delete any vertex from $V_2=\{v\in V: f(v)=2\}$ set, such as $v_4$ vertex. Since $f''(v)\neq 2$ for $\forall v \in N(v_4)$, obtained function $f''=(V_0 \cup \{v_4\}, V_1,  V_2-\{v_4\})$ does not satisfy the condition to be an RDF. According to this $\gamma_R(C_{t,r})\geq 2\frac{t}{3}+1$. For $f''$ function to be an RDF, $v_3, v_4, v_5$ vertices should be taken into $V_1$ set; $f''=(V_0-\{v_3, v_5\}, V_1\cup\{v_3, v_4, v_5\}, V_2-\{v_4\})$. Hence we get $f''(V)=2\frac{t}{3}+2$. Since $f(V)<f''(V)$ that $f''(V)\neq\gamma_R(G)$. In this case $\gamma_R(C_{t,r})\geq2\frac{t}{3}+1$.\par
Consequently $\gamma_R(C_{t,r})=2\frac{t}{3}+1$. \par
(2) $t {\equiv} 1 \text{(mod 3)}$\par
i) In order to dominate $u_1, u_2, u_3, \ldots, u_r$ , $v_1$ and $v_2$ vertices, $v_1$ vertex should be taken into $V_2$ set. To dominate $v_{t-1}$, $v_{t-2}$ and $v_{t}$ vertices, $v_{t-1}$ vertex should be taken into $V_2$ set. For the rest vertices of the graph which are not dominated, $v_t$ vertices should be taken into $V_2$ set which satisfy $t \equiv 1 \text{(mod 3)}$. Because of the $v_{t-2}$  vertex is dominated by $v_{t-3}$  vertex at the same time, taking $v_{t}$  vertex into $V_2$  set instead of $v_{t-1}$  vertex does not change the result. Then for an RDF $f$, $V_2=\lbrace v_1, v_4, v_7, \ldots, v_{t-1}\rbrace$  or  $V_2=\lbrace v_1, v_4, v_7, \ldots, v_{t}\rbrace$  and $V_1=\emptyset$ .\par
So that $f(V)=2(\frac{t-3-1}{3}+1+1)$ then we get $\gamma_R(C_{t,r})\leq 2\lceil \frac{t}{3}\rceil$. \par
Let f not be a $\gamma_R-function$ and delete any vertex from $V_2=\{v\in V: f(v)=2\}$ set, such as $v_4$ vertex. Since $f'(v)\neq 2$ for $\forall v \in N(v_4)$, obtained function $f'=(V_0 \cup \{v_4\}, \emptyset, V_2-\{v_4\})$ does not satisfy the condition to be an RDF. According to this $\gamma_R(C_{t,r})\geq 2\lceil \frac{t}{3}\rceil$. For $f'$ function to be an RDF, $v_3, v_4, v_5$ vertices should be taken into $V_1$ set; $f'=(V_0-\{v_3, v_5\}, V_1\cup\{v_3, v_4, v_5\}, V_2-\{v_4\})$. Hence we get $f'(V)=2\lceil \frac{t}{3}\rceil+1$. Since $f(V)<f'(V)$ that $f'(V)\neq\gamma_R(G)$. \par In this case $\gamma_R(C_{t,r})\geq2\lceil \frac{t}{3}\rceil$.\par
ii) In order to dominate $u_1, u_2, u_3, \ldots, u_r$ , $v_1$ and $v_2$ vertices, $v_1$ vertex should be taken into $V_2$ set. To dominate $v_{t-3}$, $v_{t-4}$ and $v_{t-2}$ vertices, $v_{t-3}$ vertex should be taken into $V_2$ set. To dominate $v_{t}$ and $v_{t-1}$ vertices, $v_{t}$ and $v_{t-1}$ themselves should be taken into $V_1$ set. For the rest vertices of the graph which are not dominated, $v_t$ vertices should be taken into $V_2$ set which satisfy $t \equiv 1 \text{(mod 3)}$. Then  $V_2=\lbrace v_1, v_4, v_7, \ldots, v_{t-3}\rbrace$  and $V_1=\lbrace v_{t-1}, v_{t}\rbrace$ . \par
Therefore $f(V)=2(\frac{t-3-1}{3}+1)+2$ then we get $\gamma_R(C_{t,r})\leq 2\lceil \frac{t}{3}\rceil$. \par
Let f not be a $\gamma_R-function$ and delete any vertex from $V_2$ set, then the result will be the same as above. So that any vertex of $V_1$ set, such as $v_t$ vertex, should be deleted. Since $f''(v)\neq 2$ for $\forall v \in N(v_t)$, obtained function $f''=(V_0 \cup \{v_t\}, V_1-\{v_t\}, V_2)$ does not satisfy the condition to be an RDF.
According to this $\gamma_R(C_{t,r})\geq 2\lceil \frac{t}{3}\rceil$. For $f''$ function to be an RDF, $v_t$ vertex should be taken into $V_2$ set; $f''=(V_0, V_1-\{v_t\}, V_2\cup\{v_t\})$. Hence we get $f''(V)=2\lceil \frac{t}{3}\rceil+1$. Since $f(V)<f''(V)$ that $f''(V)\neq\gamma_R(G)$. In this case $\gamma_R(C_{t,r})\geq2\lceil \frac{t}{3}\rceil$. \par
Consequently $\gamma_R(C_{t,r})=2\lceil\frac{t}{3}\rceil$.\par
(3) $t {\equiv} 2 \text{(mod 3)}$\par
In order to dominate $u_1, u_2, u_3, \ldots, u_r$ , $v_1$ and $v_2$ vertices, $v_1$ vertex should be taken into $V_2$ set. To dominate $v_{t-1}$, $v_{t-2}$ and $v_{t}$ vertices, $v_{t-1}$ vertex should be taken into $V_2$ set. For the rest vertices of the graph which are not dominated, $v_t$ vertices should be taken into $V_2$ set which satisfy $t \equiv 1 \text{(mod 3)}$. Then $V_2=\lbrace v_1, v_4, v_7, \ldots, v_{t-1}\rbrace$ and $V_1=\emptyset$ . \par
So that $f(V)=2(\frac{t-1-1}{3}+1)$ then we get $\gamma_R(C_{t,r})\leq 2\lceil\frac{t}{3}\rceil$. \par
Let f not be a $\gamma_R-function$ and delete any vertex from $V_2=\{v\in V: f(v)=2\}$ set, such as $v_4$ vertex. Since $f'(v)\neq 2$ for $\forall v \in N(v_4)$, obtained function $f'=(V_0 \cup \{v_4\}, \emptyset, V_2-\{v_4\})$ does not satisfy the condition to be an RDF. According to this $\gamma_R(C_{t,r})\geq 2\lceil \frac{t}{3}\rceil$. For $f'$ function to be an RDF, $v_3, v_4, v_5$ vertices should be taken into $V_1$ set; $f'=(V_0-\{v_3, v_5\}, V_1\cup\{v_3, v_4, v_5\}, V_2-\{v_4\})$. Hence we get $f'(V)=2\lceil \frac{t}{3}\rceil+1$. Since $f(V)<f'(V)$ that $f'(V)\neq\gamma_R(G)$. In this case $\gamma_R(C_{t,r})\geq2\lceil \frac{t}{3}\rceil$.\par
Consequently  $\gamma_R(C_{t,r})=2\lceil\frac{t}{3}\rceil$.\qed
\end{demo}

\begin{theorem}
For $p=n-a-b$ and $p\neq2$, let $G=DC(n, a, b)$ be a double comet graph.The roman domination number of G is equal to
\begin{equation}
    {\gamma}_R(DC(n, a, b))=
    \begin{cases}
    2(\frac{p}{3}+1) & p {\equiv} 0 \text{(mod 3)}\\
    2\lceil\frac{p}{3}\rceil & p {\equiv} 1 \text{(mod 3)}\\
    2\lceil\frac{p}{3}\rceil+1 & p {\equiv} 2 \text{(mod 3)}
    \end{cases}
\end{equation}
\end{theorem}
\begin{demo}
Roman domination number of the double comet graph is considered in three cases. Let $f=(V_0, V_1, V_2)$ be a $\gamma_R-function$ of $G$. \bigskip \par
(1) $p {\equiv} 0 \text{(mod 3)}$\par
In order to dominate $u_1, u_2, \ldots, u_a$ and $k_1, k_2$ vertices, $k_1$ vertex should be taken into $V_2$ set and similarly to dominate $v_1, v_2, \ldots, v_b$ and $k_p, k_{p-1}$ vertices, $k_p$ vertex should be taken into $V_2$ set. For the rest vertices of the graph which are not dominated $k_t$ vertices should be taken into $V_2$ set which satisfy $t\equiv1\text{(mod 3)}$ Then $V_2=\lbrace k_1, k_4, \ldots, k_{p-2}, k_p\rbrace$ and $V_1=\emptyset$.\par
So that $f(V)=2(\frac{p-2-1}{3}+1+1)$ then we get $\gamma_R(DC(n, a, b))\leq 2(\frac{p}{3}+1)$.\par

Let f not be a $\gamma_R-function$ and delete any vertex from $V_2=\{v\in V: f(v)=2\}$ set, such as $k_4$ vertex. Since $f'(v)\neq 2$ for $\forall v \in N(k_4)$, obtained function $f'=(V_0 \cup \{k_4\}, \emptyset, V_2-\{k_4\})$ does not satisfy the condition to be an RDF. According to this $\gamma_R(DC(n, a, b))\geq 2(\frac{p}{3}+1)$. For $f'$ function to be an RDF, $k_3, k_4, k_5$ vertices should be taken into $V_1$ set; $f'=(V_0-\{k_3, k_5\}, V_1\cup\{k_3, k_4, k_5\}, V_2-\{k_4\})$. Hence we get $f'(V)=2(\frac{p}{3}+1)+1$. Since $f(V)<f'(V)$ that $f'(V)\neq\gamma_R(G)$. In this case $\gamma_R(DC(n, a, b))\geq2(\frac{p}{3}+1)$.\par

Consequently  $\gamma_R(DC(n, a, b))=2(\frac{p}{3}+1)$.\par
(2) $p {\equiv} 1 \text{(mod 3)}$\par
In order to dominate $u_1, u_2, \ldots, u_a$ and $k_1, k_2$ vertices, $k_1$ vertex should be taken into $V_2$ set and similarly to dominate $v_1, v_2, \ldots, v_b$ and $k_p, k_{p-1}$ vertices, $k_p$ vertex should be taken into $V_2$ set. For the rest vertices of the graph which are not dominated $k_t$ vertices should be taken into $V_2$ set which satisfy $t\equiv1\text{(mod 3)}$ Then $V_2=\lbrace k_1, k_4, \ldots, k_{p-3}, k_p\rbrace$ and $V_1=\emptyset$.\par
So that $f(V)=2(\frac{p-3-1}{3}+1+1)$ then we get $\gamma_R(DC(n, a, b))\leq 2\lceil\frac{p}{3}\rceil$.\par
Let f not be a $\gamma_R-function$ and delete any vertex from $V_2=\{v\in V: f(v)=2\}$ set, such as $k_4$ vertex. Since $f'(v)\neq 2$ for $\forall v \in N(k_4)$, obtained function $f'=(V_0 \cup \{k_4\}, \emptyset, V_2-\{k_4\})$ does not satisfy the condition to be an RDF. According to this $\gamma_R(DC(n, a, b))\geq 2\lceil\frac{p}{3}\rceil$. For $f'$ function to be an RDF, $k_3, k_4, k_5$ vertices should be taken into $V_1$ set; $f'=(V_0-\{k_3, k_5\}, V_1\cup\{k_3, k_4, k_5\}, V_2-\{k_4\})$. Hence we get $f'(V)=2\lceil\frac{p}{3}\rceil+1$. Since $f(V)<f'(V)$ that $f'(V)\neq\gamma_R(G)$. In this case $\gamma_R(DC(n, a, b))\geq2\lceil\frac{p}{3}\rceil$.\par

Consequently  $\gamma_R(DC(n, a, b))=2\lceil\frac{p}{3}\rceil$.\par
(3) $p {\equiv} 2 \text{(mod 3)}$\par
In order to dominate $u_1, u_2, \ldots, u_a$ and $k_1, k_2$ vertices, $k_1$ vertex should be taken into $V_2$ set and similarly to dominate $v_1, v_2, \ldots, v_b$ and $k_p, k_{p-1}$ vertices, $k_p$ vertex should be taken into $V_2$ set. To dominate $k_{p-2}$ vertex, $k_{p-2}$ vertex itself should be taken into $V_1$ set. For the rest vertices of the graph which are not dominated $k_t$ vertices should be taken into $V_2$ set which satisfy $t\equiv1\text{(mod 3)}$ Then $V_2=\lbrace k_1, k_4, \ldots, k_{p-4}, k_p\rbrace$ and $V_1=\lbrace k_{p-2}\rbrace$. \par
So that $f(V)=2(\frac{p-4-1}{3}+1+1)+1$ then we get $\gamma_R(DC(n, a, b))\leq 2\lceil\frac{p}{3}\rceil+1$.\par

Let f not be a $\gamma_R-function$ and by deleting $k_{p-2}$ vertex from $V_1=\{v\in V: f(v)=1\}$ set, let $V_1=\emptyset$. Since $f'(v)\neq 2$ for $\forall v \in N(k_{p-2})$, obtained function $f'=(V_0 \cup \{k_{p-2}\}, \emptyset, V_2)$ does not satisfy the condition to be an RDF. According to this $\gamma_R(DC(n, a, b))\geq 2\lceil\frac{p}{3}\rceil+1$. For $f'$ function to be an RDF, $k_{p-2}$ vertex should be taken into $V_2$ set; $f'=(V_0, \emptyset, V_2\cup\{k_{p-2}\})$. Hence we get $f'(V)=2\lceil\frac{p}{3}\rceil+2$. Since $f(V)<f'(V)$ that $f'(V)\neq\gamma_R(G)$. In this case $\gamma_R(DC(n, a, b))\geq 2\lceil\frac{p}{3}\rceil+1$.\par
Let f not be a $\gamma_R-function$ and delete any vertex from $V_2=\{v\in V: f(v)=2\}$ set, such as $k_4$ vertex. Since $f''(v)\neq 2$ for $\forall v \in N(k_4)$, obtained function $f''=(V_0 \cup \{k_4\}, V_1, V_2-\{k_4\})$ does not satisfy the condition to be an RDF. According to this $\gamma_R(DC(n, a, b))\geq 2\lceil\frac{p}{3}\rceil+1$. For $f''$ function to be an RDF, $k_3, k_4, k_5$ vertices should be taken into $V_1$ set; $f''=(V_0-\{k_3, k_5\}, V_1\cup\{k_3, k_4, k_5\}, V_2-\{k_4\})$. Hence we get $f''(V)=2\lceil\frac{p}{3}\rceil+2$. Since $f(V)<f''(V)$ that $f''(V)\neq\gamma_R(G)$. In this case $\gamma_R(DC(n, a, b))\geq 2\lceil\frac{p}{3}\rceil+1$.\par
Consequently  $\gamma_R(DC(n, a, b))=2\lceil\frac{p}{3}\rceil+1$.\qed
\end{demo}
\begin{theorem}
Let $G=P_n^+$ be a comb graph. The roman domination number of $G$ is equal to
\begin{equation}
{\gamma}_R(P_n^+)=
\begin{cases}
4\frac{n}{3} & n  {\equiv}0  \text{(mod 3)}\\
4\lfloor\frac{n}{3}\rfloor+2 & n  {\equiv}1  \text{(mod 3)}\\
4\lceil \frac{n}{3} \rceil-1 &   n  {\equiv}2  \text{(mod 3)}\\
\end{cases}
\end{equation}
\end{theorem}
\begin{demo}
Roman domination number of the comb graph is considered in three cases.  Let $f=(V_0, V_1, V_2)$ be a $\gamma_R-function$ of $G$. \bigskip\par
(1) $n {\equiv} 0 \text{(mod 3)}$\par
In order to dominate $v_{t-1}, v_t, v_{t+1}$ and $v_t^+$ vertices, $v_t$ vertices should be taken into $V_2$ set which satisfy $t \equiv 1 \text {(mod 3)}$.  For the rest vertices of the graph which are not dominated, $v_{t-1}^+$  and $v_{t+1}^+$ vertices should be taken into $V_1$ set which satisfy $t \equiv 1 \text{(mod 3)}$. Then $V_2=\lbrace v_1, v_4, v_7, \ldots, v_{n-5}, v_{n-2} \rbrace$ and $V_1=\lbrace v_{0}^+, v_{2}^+, v_{3}^+, v_{5}^+, \ldots, v_{n-3}^+, v_{n-1}^+\rbrace$ . \par
So that $f(V)=(\frac{n-3-0}{3}+1+\frac{n-1-2}{3}+1)+2(\frac{n-2-1}{3}+1)$ then we get $\gamma_R(P_n^+)\leq 4\frac{n}{3}$.\par

Let f not be a $\gamma_R-function$ and delete any vertex from $V_1=\{v\in V: f(v)=1\}$ set, such as $v_0^+$ vertex. Since $f'(v)\neq 2$ for $\forall v \in N(v_0^+)$, obtained function $f'=(V_0 \cup \{v_0^+\}, V_1-\{v_0^+\}, V_2)$ does not satisfy the condition to be an RDF. According to this $\gamma_R(P_n^+)\geq 4\frac{n}{3}$. For $f'$ function to be an RDF, $v_0^+$ vertex should be taken into $V_2$ set; $f'=(V_0, V_1-\{v_0^+\}, V_2\cup\{v_0^+\})$. Hence we get $f'(V)=4\frac{n}{3}+1$. Since $f(V)<f'(V)$ that $f'(V)\neq\gamma_R(G)$. In this case $\gamma_R(P_n^+)\geq 4\frac{n}{3}$.\par

Let f not be a $\gamma_R-function$ and delete any vertex from $V_2=\{v\in V: f(v)=2\}$ set, such as $v_1$ vertex. Since $f''(v)\neq 2$ for $\forall v \in N(v_1)$, obtained function $f''=(V_0 \cup \{v_1\}, V_1, V_2-\{v_1\})$ does not satisfy the condition to be an RDF. According to this $\gamma_R(P_n^+)\geq 4\frac{n}{3}$. For $f''$ function to be an RDF, $v_0, v_1, v_2$ and $v_1^+$ vertices should be taken into $V_1$ set; $f''=(V_0-\{v_0, v_2, v_1^+\}, V_1\cup\{v_0, v_1, v_2, v_1^+\}, V_2-\{v_1\})$. Hence we get $f''(V)=4\frac{n}{3}+2$. Since $f(V)<f''(V)$ that $f''(V)\neq\gamma_R(G)$. In this case $\gamma_R(P_n^+)\geq 4\frac{n}{3}$.\par

Consequently  $\gamma_R(P_n^+)=4\frac{n}{3}$.
\bigskip\par
(2) $n{\equiv}1\text{(mod 3)}$\par
i) In order to dominate $v_{t-1}, v_t, v_{t+1}$ and $v_t^+$  vertices, $v_t$  vertices should be taken into $V_2$  set which satisfy $t{\equiv}1\text{(mod 3)}$ . To dominate $v_{n-1}$  and $v_{n-1}^+$  vertices,  $v_{n-1}$  or $v_{n-1}^+$ vertex should be taken into $V_2$  set. For the rest vertices of the graph which are not dominated,  $v_{t-1}^+$ and $v_{t+1}^+$ vertices should be taken into  $V_1$ set which satisfy $t{\equiv}1\text{(mod 3)}$ . Then $V_2=\lbrace v_1, v_4, v_7, \ldots, v_{n-6}, v_{n-3}, v_{n-1}\rbrace$  or $V_2=\lbrace v_1, v_4, v_7, \ldots, v_{n-6}, v_{n-3}, v_{n-1}^+ \rbrace$  and $V_1=\lbrace v_0^+, v_2^+, v_3^+, v_5^+, \ldots, v_{n-4}^+, v_{n-2}^+\rbrace$. \par
Thus, $f(V)=(\frac{n-4-0}{3}+1+\frac{n-2-2}{3}+1)+2(\frac{n-3-1}{3}+1+1)$  then we get $\gamma_R(P_n^+)\leq 4\lfloor\frac{n}{3}\rfloor+2$.\par

Let f not be a $\gamma_R-function$ and delete any vertex from $V_1=\{v\in V: f(v)=1\}$ set, such as $v_0^+$ vertex. Since $f'(v)\neq 2$ for $\forall v \in N(v_0^+)$, obtained function $f'=(V_0 \cup \{v_0^+\}, V_1-\{v_0^+\}, V_2)$ does not satisfy the condition to be an RDF. According to this $\gamma_R(P_n^+)\geq  4\lfloor\frac{n}{3}\rfloor+2$. For $f'$ function to be an RDF, $v_0^+$ vertex should be taken into $V_2$ set; $f'=(V_0, V_1-\{v_0^+\}, V_2\cup\{v_0^+\})$. Hence we get $f'(V)=4\lfloor\frac{n}{3}\rfloor+3$. Since $f(V)<f'(V)$ that $f'(V)\neq\gamma_R(G)$. In this case $\gamma_R(P_n^+)\geq 4\lfloor\frac{n}{3}\rfloor+2$.\par
Let f not be a $\gamma_R-function$ and delete any vertex from $V_2=\{v\in V: f(v)=2\}$ set, such as $v_1$ vertex. Since $f''(v)\neq 2$ for $\forall v \in N(v_1)$, obtained function $f''=(V_0 \cup \{v_1\}, V_1, V_2-\{v_1\})$ does not satisfy the condition to be an RDF. According to this $\gamma_R(P_n^+)\geq  4\lfloor\frac{n}{3}\rfloor+2$. For $f''$ function to be an RDF, $v_0, v_1, v_2$ and $v_1^+$ vertices should be taken into $V_1$ set; $f''=(V_0-\{v_0, v_2, v_1^+\}, V_1\cup\{v_0, v_1, v_2, v_1^+\}, V_2-\{v_1\})$. Hence we get $f''(V)=4\lfloor\frac{n}{3}\rfloor+4$. Since $f(V)<f''(V)$ that $f''(V)\neq\gamma_R(G)$. In this case $\gamma_R(P_n^+)\geq 4\lfloor\frac{n}{3}\rfloor+2$.\par
ii) In order to dominate $v_{t-1}, v_t, v_{t+1}$ and $v_t^+$  vertices, $v_t$  vertices should be taken into $V_2$  set which satisfy $t{\equiv}1\text{(mod 3)}$ . To dominate $v_{n-1}$  and $v_{n-1}^+$  vertices,  $v_{n-1}$  and $v_{n-1}^+$ vertices should be taken into $V_1$  set. For the rest vertices of the graph which are not dominated,  $v_{t-1}^+$ and $v_{t+1}^+$ vertices should be taken into  $V_1$ set which satisfy $t{\equiv}1\text{(mod 3)}$ . Then \\$V_2=\lbrace v_1, v_4, v_7, \ldots, v_{n-6}, v_{n-3}\rbrace$  and $V_1=\lbrace v_0^+, v_2^+, v_3^+, v_5^+, \ldots, v_{n-4}^+, v_{n-2}^+, v_{n-1}^+, v_{n-1}\rbrace$. \par
So that $f(V)=(\frac{n-4-0}{3}+1+\frac{n-2-2}{3}+1+2)+2(\frac{n-3-1}{3}+1)$ then we get $\gamma_R(P_n^+)\leq 4\lfloor\frac{n}{3}\rfloor+2$.\par

Let f not be a $\gamma_R-function$ and delete any vertex from $V_1=\{v\in V: f(v)=1\}$ set, such as $v_0^+$ vertex. Since $f'(v)\neq 2$ for $\forall v \in N(v_0^+)$, obtained function $f'=(V_0 \cup \{v_0^+\}, V_1-\{v_0^+\}, V_2)$ does not satisfy the condition to be an RDF. According to this $\gamma_R(P_n^+)\geq  4\lfloor\frac{n}{3}\rfloor+2$. For $f'$ function to be an RDF, $v_0^+$ vertex should be taken into $V_2$ set; $f'=(V_0, V_1-\{v_0^+\}, V_2\cup\{v_0^+\})$. Hence we get $f'(V)=4\lfloor\frac{n}{3}\rfloor+3$. Since $f(V)<f'(V)$ that $f'(V)\neq\gamma_R(G)$. In this case $\gamma_R(P_n^+)\geq 4\lfloor\frac{n}{3}\rfloor+2$.\par
Let f not be a $\gamma_R-function$ and delete any vertex from $V_2=\{v\in V: f(v)=2\}$ set, such as $v_1$ vertex. Since $f''(v)\neq 2$ for $\forall v \in N(v_1)$, obtained function $f''=(V_0 \cup \{v_1\}, V_1, V_2-\{v_1\})$ does not satisfy the condition to be an RDF. According to this $\gamma_R(P_n^+)\geq  4\lfloor\frac{n}{3}\rfloor+2$. For $f''$ function to be an RDF, $v_0, v_1, v_2$ and $v_1^+$ vertices should be taken into $V_1$ set; $f''=(V_0-\{v_0, v_2, v_1^+\}, V_1\cup\{v_0, v_1, v_2, v_1^+\}, V_2-\{v_1\})$. Hence we get $f''(V)=4\lfloor\frac{n}{3}\rfloor+4$. Since $f(V)<f''(V)$ that $f''(V)\neq\gamma_R(G)$. In this case $\gamma_R(P_n^+)\geq 4\lfloor\frac{n}{3}\rfloor+2$.\par

Consequently, we could say that $\gamma_R(P_n^+)=4\lfloor\frac{n}{3}\rfloor+2$.\par
(3) $n{\equiv}2 \text{(mod 3)}$\par
i) In order to dominate $v_{t-1}, v_t, v_{t+1}$ and $v_t^+$  vertices, $v_t$  vertices should be taken into $V_2$  set which satisfy $t\equiv1\text{(mod 3)}$ . For the rest vertices of the graph which are not dominated, $v_{t-1}^+$  and $v_{t+1}^+$ vertices should be taken into $V_1$  set which satisfy $t\equiv1\text{(mod 3)}$ . Then $V_2=\lbrace v_1, v_4, v_7, \ldots, v_{n-4}, v_{n-1}\rbrace$  and $V_1=\lbrace v_0^+, v_2^+, v_3^+, v_5^+, \ldots, v_{n-3}^+, v_{n-2}^+\rbrace$ .\par
So that $f(V)=(\frac{n-2-0}{3}+1+\frac{n-3-2}{3}+1)+2(\frac{n-1-1}{3}+1)$ then we get $\gamma_R(P_n^+)\leq 4\lceil\frac{n}{3}\rceil-1$.\par

Let f not be a $\gamma_R-function$ and delete any vertex from $V_1=\{v\in V: f(v)=1\}$ set, such as $v_0^+$ vertex. Since $f'(v)\neq 2$ for $\forall v \in N(v_0^+)$, obtained function $f'=(V_0 \cup \{v_0^+\}, V_1-\{v_0^+\}, V_2)$ does not satisfy the condition to be an RDF. According to this $\gamma_R(P_n^+)\geq  4\lceil\frac{n}{3}\rceil-1$. For $f'$ function to be an RDF, $v_0^+$ vertex should be taken into $V_2$ set; $f'=(V_0, V_1-\{v_0^+\}, V_2\cup\{v_0^+\})$. Hence we get $f'(V)=4\lceil\frac{n}{3}\rceil$. Since $f(V)<f'(V)$ that $f'(V)\neq\gamma_R(G)$. In this case $\gamma_R(P_n^+)\geq 4\lceil\frac{n}{3}\rceil-1$.\par
Let f not be a $\gamma_R-function$ and delete any vertex from $V_2=\{v\in V: f(v)=2\}$ set, such as $v_1$ vertex. Since $f''(v)\neq 2$ for $\forall v \in N(v_1)$, obtained function $f''=(V_0 \cup \{v_1\}, V_1, V_2-\{v_1\})$ does not satisfy the condition to be an RDF. According to this $\gamma_R(P_n^+)\geq  4\lceil\frac{n}{3}\rceil-1$. For $f''$ function to be an RDF, $v_0, v_1, v_2$ and $v_1^+$ vertices should be taken into $V_1$ set; $f''=(V_0-\{v_0, v_2, v_1^+\}, V_1\cup\{v_0, v_1, v_2, v_1^+\}, V_2-\{v_1\})$. Hence we get $f''(V)=4\lceil\frac{n}{3}\rceil+1$. Since $f(V)<f''(V)$ that $f''(V)\neq\gamma_R(G)$. In this case $\gamma_R(P_n^+)\geq 4\lceil\frac{n}{3}\rceil-1$.\par
ii) To dominate $v_{n-2}, v_{n-1}$  and $v_{n-2}^+$  vertices,  $v_{n-2}$ vertex should be taken into $V_2$  set. To dominate $v_{n-1}^+$ vertex, $v_{n-1}^+$ vertex itself should be taken into $V_1$  set. For the rest vertices of the graph which are not dominated, in order to dominate $v_{t-1}, v_t, v_{t+1}$ and $v_t^+$  vertices, $v_t$  vertices should be taken into $V_2$  set , and  $v_{t-1}^+$ and $v_{t+1}^+$ vertices should be taken into  $V_1$ set which satisfy $t{\equiv}1\text{(mod 3)}$ . Then $V_2=\lbrace v_1, v_4, v_7, \ldots, v_{n-4}, v_{n-2}\rbrace$  and $V_1=\lbrace v_0^+, v_2^+,v _3^+, v_5^+, \ldots, v_{n-3}^+, v_{n-1}^+\rbrace$. \par
Therefore $f(V)=(\frac{n-5-0}{3}+1+\frac{n-3-2}{3}+1+1)+2(\frac{n-4-1}{3}+1+1)$ then we get $\gamma_R(P_n^+)\leq 4\lceil\frac{n}{3}\rceil-1$.\par
Let f not be a $\gamma_R-function$ and delete any vertex from $V_1=\{v\in V: f(v)=1\}$ set, such as $v_0^+$ vertex. Since $f'(v)\neq 2$ for $\forall v \in N(v_0^+)$, obtained function $f'=(V_0 \cup \{v_0^+\}, V_1-\{v_0^+\}, V_2)$ does not satisfy the condition to be an RDF. According to this $\gamma_R(P_n^+)\geq  4\lceil\frac{n}{3}\rceil-1$. For $f'$ function to be an RDF, $v_0^+$ vertex should be taken into $V_2$ set; $f'=(V_0, V_1-\{v_0^+\}, V_2\cup\{v_0^+\})$. Hence we get $f'(V)=4\lceil\frac{n}{3}\rceil$. Since $f(V)<f'(V)$ that $f'(V)\neq\gamma_R(G)$. In this case $\gamma_R(P_n^+)\geq 4\lceil\frac{n}{3}\rceil-1$.\par
Let f not be a $\gamma_R-function$ and delete any vertex from $V_2=\{v\in V: f(v)=2\}$ set, such as $v_1$ vertex. Since $f''(v)\neq 2$ for $\forall v \in N(v_1)$, obtained function $f''=(V_0 \cup \{v_1\}, V_1, V_2-\{v_1\})$ does not satisfy the condition to be an RDF. According to this $\gamma_R(P_n^+)\geq  4\lceil\frac{n}{3}\rceil-1$. For $f''$ function to be an RDF, $v_0, v_1, v_2$ and $v_1^+$ vertices should be taken into $V_1$ set; $f''=(V_0-\{v_0, v_2, v_1^+\}, V_1\cup\{v_0, v_1, v_2, v_1^+\}, V_2-\{v_1\})$. Hence we get $f''(V)=4\lceil\frac{n}{3}\rceil+1$. Since $f(V)<f''(V)$ that $f''(V)\neq\gamma_R(G)$. In this case $\gamma_R(P_n^+)\geq 4\lceil\frac{n}{3}\rceil-1$.\par
Consequently, we had $\gamma_R(P_n^+)=4\lceil\frac{n}{3}\rceil-1$.\qed
\end{demo}

\nocite{7}
\nocite{8}
\nocite{9}
\nocite{10}
\nocite{11}
\nocite{12}

\end{document}